\documentclass[showpacs,prl,english,french,twocolumn]{revtex4}
\usepackage[latin9]{inputenc}
\usepackage{graphicx}
\usepackage{esint}
\usepackage{amssymb}
\usepackage{epsfig,epsf}
\usepackage{pstricks}
\usepackage{pstricks-add}
\usepackage{pst-plot}
\usepackage{babel}
\usepackage[babel=true,kerning=true]{microtype}
\usepackage{soul}

\RequirePackage{color}
\definecolor{MyDarkGreen}{rgb}{0.02,0.60,0.06}

\begin{document}

\title{Optics near an hyperbolic defect}

\author{S\'ebastien Fumeron, Bertrand Berche}

\affiliation{Universite de Lorraine - UMR CNRS 7198
 BP 70239,  54506 Vand\oe uvre les Nancy, France}
 
\author{Fernando Santos}
\altaffiliation[On leave from:] {Departamento de Matem\'atica  Universidade Federal 
de Pernambuco, 50670-901, Recife, PE, Brazil}

\affiliation{Wolfson Centre for Mathematical Biology, Mathematical Institute, University of Oxford, OX1 3LB Oxford, U.K.}

\author{Erms Pereira}
\affiliation{Escola Polit\'ecnica de Pernambuco, Universidade de Pernambuco, Rua
Benfica, 455, 50720-001, Recife, PE, Brazil}

\author{Fernando Moraes}

\altaffiliation[On leave from: ]{Departamento de F\'{\i} sica, CCEN, Universidade Federal da Para\'\i ba, Caixa Postal 5008, 58051-900 , Jo\~ao Pessoa, PB, Brazil }

\selectlanguage{english}%

\affiliation{Departamento de F\'\i sica and Departamento de Matem\'atica, Universidade
Federal de Pernambuco, 
 50670-901 Recife, PE, Brazil}

\begin{abstract}
We examine the properties of a new family of defects called hyperbolic disclinations, and discuss their possible use for the design of perfect optical absorbers. In hyperbolic metamaterials, the ratio of ordinary and extraordinary permittivities is negative, which leads to an effective metric of Kleinian signature (two timelike coordinates). Considering a disclination in the hyperbolic nematic host matrix, we show that the timelike geodesics are Poinsot spirals, i.e. whatever the impact parameter of an incident light beam, it is confined and whirls about the defect core. The trapping effect does not require light to be coherent. This property also remains in the wave formalism, which may be the sign for many potential applications.
\end{abstract}


\maketitle


Since the pioneering works of Veselago \cite{veselago}, media exibiting negative permittivities (better known as metamaterials) led to an unprecedented control of light. Cloaking the electromagnetic field, imaging objects beyond the diffraction limit (perfect lens) or manipulating the near field are only few examples of the new possibilities offered by these artificial media \cite{pendry}. Besides these technological applications, research on metamaterials also benefits to fundamental physics: quite unexpectedly, it provides new tools to investigate optical analogs of black holes \cite{philbin, genov}. Recently, the advent of hyperbolic (or indefinite) metamaterials represented another step forward in light control (for a review, see Ref. \cite{Poddubny}) as well as in testing high energy physics (metric signature transitions \cite{smolyanov}). Such media exibit effective permittivities of opposite signs and can be made from a host nematic liquid crystal doped with coated core-shell nanospheres \cite{Pawlik}. 

In this work, we will investigate the propagation of light in an hyperbolic liquid crystal endowed with a linear defect called disclination. The hyperbolic behavior modifies the regular conical geometry around the disclination, leading to a new family of topological defects: the hyperbolic disclination. The study of null geodesics shows that the effective metric induced by the defect has a dramatic influence on light propagation: rays are totally trapped by the disclination, regardless of the impact parameter. Then, as the geometry of hyperbolic disclination was never investigated, some of its general properties such as the group of associated spacetime isometries are examined. Finally, the standpoint of wave optics is used to study hyperbolic disclinations and their relevance for the design of optical absorbers is briefly discussed.

In practice, the hyperbolic medium can be made of an host nematic liquid crystal containing an uniform distribution of coated core-shell spheres (polaritonic core and semiconductor shell) and submitted to an external electric field \cite{Pawlik}. For a planar anchoring, the director field is orthoradial, such that the uniaxial permittivity tensor is given by 
\begin{equation}
\overline{\overline{\varepsilon}}=\varepsilon_o \;\hat{r}\otimes\hat{r}+\varepsilon_e \left(\frac{1}{r^2}\hat{\phi}\otimes\hat{\phi}\right)+\varepsilon_o\hat{z}\otimes\hat{z}
\end{equation}
\noindent where $\varepsilon_o$ denotes the ordinary permittivity, $\varepsilon_e$ denotes the extraordinary permittivity and $\hat{r}, \hat{\phi}, \hat{z}$ stand respectively for the three unit vectors of the cylindrical coordinate basis. Uniaxial crystals support two eigenmodes: the ordinary mode that experiences the medium as an isotropic dielectric with permittivity $\varepsilon_o$, and the extraordinary mode, that experiences an anisotropic refractive index. Extraordinary rays obey an hyperbolic dispersion relation \cite{Pawlik}:
\begin{equation}
\frac{k_{\perp}^2}{\varepsilon_o}+\frac{k_{\shortparallel}^2}{\varepsilon_e}=\frac{\omega^2}{c^2}
\end{equation}
Depending on incident angle of light, either $\varepsilon_o$ or $\varepsilon_e$ can be negative (magnetic permeability remains equal to unity). In the remainder, we will restrict to the extraordinary light, which is polarized in the plane defined by the wavevector $\bold{k}$ and the director \cite{oswald}. The director field exibits cylindrical symmetry, as happens for nematics confined inside a capillary tube (axis $z$), and light beams are shot in planes of constant height. Then, using the formulation of Fermat principle in terms of light geodesics in a Riemannian curved geometry, the metric experienced by light in the vicinity of a disclination (with unit topological charge) is given by \cite{satiro}: 

\begin{equation}
ds^{2}=-c^2dt^2+\varepsilon_o dr^{2}+\varepsilon_e r^{2} d\phi^2+\varepsilon_o dz^{2}
\end{equation}
Such a configuration is carried out with hometropic anchoring at the boundaries. In the following, since we are interested in the extraordinary ray propagating in $z={\rm const.}$ planes, we will omit the $z$ component in the effective metric from now on. Assuming for example that $\varepsilon_o \geqslant 0$ and thus $\varepsilon_e\leqslant 0$, one can rescale the radial coordinate by $\rho=\sqrt{\varepsilon_o}r$ such that 
\begin{eqnarray}
ds^{2}=-c^2dt^2+d\rho^{2}+\alpha^2 \rho^{2}d\phi^2 \label{eq-metrics}
\end{eqnarray}
The third term being negative, this metrics corresponds to a disclination with a defect parameter $\alpha^2=\varepsilon_e/\varepsilon_o=\left(i \gamma\right)^2\leqslant 0$, or equivalently to a disclination with hyperbolic (imaginary) deficit angle \textit{i}$\times 2\pi\gamma $. Note also that this metrics is not valid at $\rho=0$ and an ultraviolet cut-off is required at small radii because of the core structure of the topological defect~\cite{deGennes}. Eq.~(\ref{eq-metrics}) can be recast into a canonical form by the coordinate transformation $X=\rho\cosh \gamma\phi$ and $Y=\rho\sinh \gamma\phi$:
\begin{equation}
ds^{2}=-c^2dt^2+ dX^{2}-dY^2 \label{hypo3}
\end{equation}
The unusual signature $\left(-,+,-\right)$ is known as a Kleinian signature (after pioneering works by Felix Klein) and (\ref{hypo3}) corresponds to an ultrahyperbolic or Pl\"ucker geometry \cite{Gibbons}. The coordinate $Y$ (or equivalently $\phi$) behaves as a "pseudo time coordinate". Only few theoretical investigations for such geometry have been made, and to the best of our knowledge, these were limited to the context of quantum gravity \cite{Alty,Visser}. The metric signature transition between ordinary Minkowski space-time and an effectively Kleinian space-time has been discussed from the point of view of metamaterials in \cite{smolyanov,gomez}.


In the geometrical optics approximation, the trajectories followed by light are the null geodesics of the hyperbolic metric and they are determined by
\begin{equation}
\frac{d^2 x^{\mu}}{d\lambda^2}+\Gamma^{\mu}_{\nu\sigma}\frac{d x^{\nu}}{d\lambda}\frac{d x^{\sigma}}{d\lambda}=0 \label{geod}
\end{equation} 
where the $\Gamma^{\mu}_{\nu\sigma}$ are the affine connections (here the Christoffel symbols of second kind) and $\lambda$ is an affine parameter. Only three Christoffel symbols do not vanish:
$\Gamma^{\rho}_{\phi\phi}=\gamma^2 \rho$, $\Gamma^{\phi}_{\rho\phi}=\Gamma^{\phi}_{\phi\rho}=\frac{1}{\rho}$.
Susbtituting into the geodesic equations (\ref{geod}), one is left with three independent equations:
\begin{eqnarray}
\frac{dt}{d\lambda}=\kappa \\
\ddot{\phi}+\frac{2}{\rho} \dot{\rho}\;\dot{\phi}&=&0 \label{g1} \\
\ddot{\rho}+\gamma^2 \rho \;\dot{\phi}^2&=&0 \label{g2}
\end{eqnarray}
where $\kappa$ is a real number and the dot notation stands for a derivative with respect to $\lambda$. The second equation is equivalent to the conservation of angular momentum for a unit mass and integrates straightforwardly into:
$\rho^2 \dot{\phi}=L .$ 
Instead of integrating (\ref{g2}), one uses directly a first integral of motion, obtained from the line element. 
As light follows null geodesics of effective spacetime, it comes out
\begin{equation}
0=g_{\mu\nu}\dot{x}^{\mu}\dot{x}^{\nu}=-\kappa^2+\dot{\rho}^2-\gamma^2\rho^2\dot{\phi}^2
\end{equation}
Substitution into (\ref{g2}) leads to:
\begin{equation}
\frac{1}{2}\dot{\rho}^2-\gamma^2 \frac{L^2}{2 \rho^2} = \frac{\kappa^2}{2}=\tilde{E}\geqslant 0 \label{g4}
\end{equation}
This equation can be interpreted as the total energy of a unit mass particle moving radially in an effective power-law potential. Thus, the hyperbolic defect generates an attractive force towards the defect:
\begin{equation}
F_d=-\gamma^2 \frac{L^2}{\rho^3},
\end{equation}
which can be understood a centripetal force in three dimensions. Application of Bertrand's theorem shows that light paths are generally not closed for such a geometry. 

\begin{figure*}
\noindent\includegraphics[height=7cm]{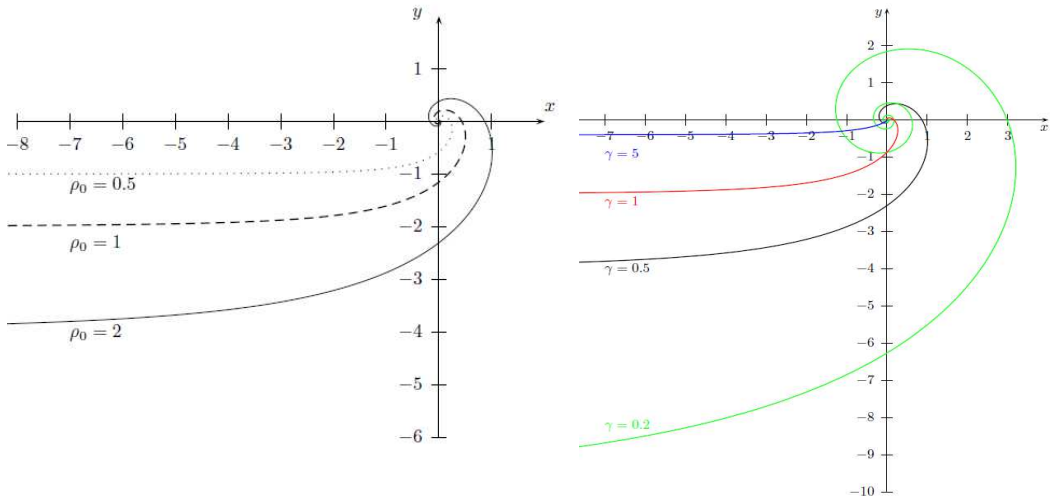}
\caption{\textit{Left}: Variations of the Poinsot spiral (positive sign) with respect to parameter $\rho_0$ ($\gamma=0.5$). \textit{Right}: Variations of the Poinsot spiral (positive sign) with respect to whirling parameter $\gamma$ ($\rho_0=2$). Solutions corresponding to the minus sign are deduced by a rotation of 180\degre.}\label{poinsot1}
\end{figure*}

\noindent Assuming that $\tilde{E}\neq 0$, solutions of geodesic equations are obtained as Poinsot spirals according to:
\begin{eqnarray}
\rho\left(\phi\right)&=& -\frac{\rho_0}{\sinh \gamma\phi} \;\;\;\text{if $\phi\geqslant 0$} \label{poins1} \\
&=&\frac{\rho_0}{\sinh \gamma\phi} \;\;\;\;\;\;\text{if $\phi\leqslant 0$} \label{poins2}
\end{eqnarray}
where $\rho_0=1/\sqrt{\frac{2 \tilde{E}}{\gamma^2 L^2}}$. In Fig.\ref{poinsot1} a family of curves corresponding to (\ref{poins1}) is plotted. Any ray incoming onto the defect with impact parameter  $\rho_0/\gamma$ spirals around the asymptotic point $\rho=0$. The smaller the value of the parameter $\gamma$, the more important the effect, hence $1/\gamma$ can be understood as the whirling strength (or vorticity) of the defect.

Due to the hyperbolic geometry, the group of isometries preserving the Pl\"ucker line element differs from the usual Poincaré group. Besides usual translations and $t-X$ boost, one finds two new kinds of transformations. The first one is given by

\begin{eqnarray}
ct^{\prime}&=&ct \label{nb1} \\
X^{\prime}&=&\Gamma\left(X-\frac{v}{c}Y\right) \label{nb2} \\
Y^{\prime}&=&\Gamma\left(-\frac{v}{c}X+Y\right) \label{nb3} 
\end{eqnarray}
where $\Gamma=(1-v^2/c^2)^{-1/2}$ is analog to the usual Lorentz factor, and which implies that the paramater $v$ is bounded by $c$. This is similar to a space-time rotation (or boost) between $Y$ and $X$, the rapidity being given by $\Psi=\text{arctanh}(v/c)$. The second kind of transformation is given by:
\begin{eqnarray}
ct^{\prime}&=&\tilde{\Gamma} \left(ct+\frac{u}{c}Y\right) \label{nr1} \\
X^{\prime}&=&X \label{nr2} \\
Y^{\prime}&=&\tilde{\Gamma} \left(-u t+Y\right) \label{nr3} 
\end{eqnarray}
where 
$\tilde{\Gamma}=(1+v^2/c^2)^{-1/2}$.
This is a rotation between two timelike coordinates $t$ and $Y$, the angle being defined by $\Theta=\arctan(u/c)$. In this case, the requirement for $\tilde{\Gamma}$ to be defined does set any limit on $u$. For Kleinian spacetimes, the set of transformations leaving (\ref{hypo3}) unchanged is a 6-parameter group which includes five kinds of transformations: the three space and time translations, one regular boost (between $t$ and space coordinate $X$), but there appear also one pseudo-boost (between pseudo-time $Y$ and $X$) and one time rotation (between time $t$ and pseudo-time $Y$).

$Y$ being looked upon as a pseudo-time coordinate does not threaten causality, as it arises from the effective geometry treatment of the material. However, classical systems are now extensively used to investigate their high-energy counterparts (for classical optics, see for example \cite{Eilenberger,Philbin}), such that the question of causality cannot be circumvented when considering true metrics with Kleinian signature. In particular, do the null geodesics (\ref{poins1})-(\ref{poins2}) preserve causality ? Causality is preserved when the sign of time interval between two events is preserved under the elements of the Poincar\'e group. In the case of Pl\"ucker geometry, this is true for regular boosts, space rotations and boosts, and it is obviously verified for pseudo-boosts. Considering the time rotation (\ref{nr1})-(\ref{nr3}), time intervals are related each other by:
\begin{eqnarray}
d t^{\prime}&=&\tilde{\Gamma}d t \left(1+\frac{u}{c^2}\frac{d Y}{dt}\right) \label{nr1}
\end{eqnarray}
Hence, causality is preserved independently of values taken by $u$ (the unbounded parameter) for events such that when $dY=0$, that is
$0=\sinh\left( \gamma\phi\right) d\rho + \rho\gamma\cosh \left(\gamma\phi\right) d\phi $
which is satisfied by both solutions (\ref{poins1})-(\ref{poins2}). In other words, causality is preserved for all events on null geodesics in the Pl\"ucker geometry.

Let us now examine the problem of light propagation in the framework of wave optics. In the scalar wave approximation, light wave dynamics is ruled by the covariant d'Alembert equation
\cite{Uzan} 
\begin{equation}
\nabla_{\mu}\nabla^{\mu}\Phi=\frac{1}{\sqrt{-g}}\partial_{\mu}\left(\sqrt{-g}g^{\mu\nu}\partial_{\nu}\Phi\right)=0 \label{eq:alembert}
\end{equation}
where $\Phi$ is the wave function, $g^{\mu\nu}$ is the metric in contravariant form and $g=\det\left(g_{\mu\nu}\right)$. Assuming that the dependency with respect to $t$ is harmonic and expressing the operator in terms of variables $X$ and $Y$ (wavevector has a null z-component), this becomes:
\begin{equation}
\left(\frac{\partial^2 \Phi}{\partial Y^2}-\frac{\partial^2 \Phi}{\partial X^2}\right)-\frac{\omega^2}{c^2}\Phi 
=0 \label{KG}
\end{equation}
where $\omega$ is the angular frequency. The 4-wavevector can be written in contravariant components as $K^{\mu}=(\omega/c,k,\Omega/c,0)$ and solutions of (\ref{KG}) are linear combinations of plane waves of the form:
\begin{eqnarray}
\Phi(X,Y,t)&=&\Phi_0\exp\left(i k_{\mu}x^{\mu}\right) \\
&=&\Phi_0\exp\left(i\left[k X-(\Omega/c) Y-\omega t\right]\right) \label{PWH}
\end{eqnarray}
which obey the dispersion relation 
\begin{equation}
\omega^2=c^2k^2-\Omega^2
\end{equation}
This dispersion relation presents a forbidden band as occurs for low-frequency phonon polaritons \cite{Kittel}. 

\begin{figure*}
\noindent\includegraphics[width=15cm]{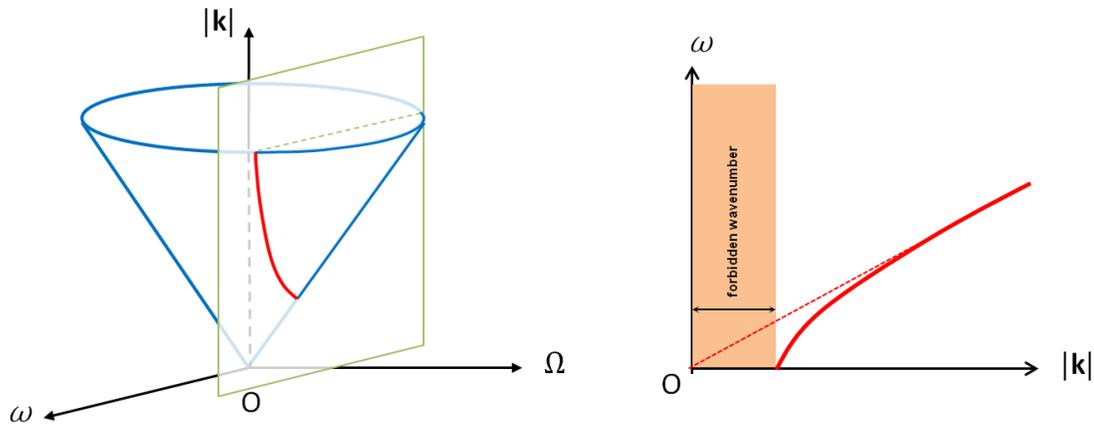}
\caption{\textit{Left}: Conical structure of the hyperbolic dispersion relation. \textit{Right}: Plot of the dispersion relation showing a zone of forbidden wavenumbers.}\label{disp}
\end{figure*}

In cylindrical coordinates (used by an external observer), d'Alembert equation writes as
\begin{equation}
-\partial^2_{t}\Phi+\frac{1}{\rho}\partial_{\rho}\left(\rho\partial_{\rho}\Phi\right)-\frac{1}{\gamma^2\rho^2}\partial^2_{\phi}\Phi=0 \label{eq:alembert1}
\end{equation}
Performing a Jacobi-Anger expansion from (\ref{PWH}) leads to seek solutions of the form:
\begin{equation}
\Phi\left(t,\rho,\phi\right)=e^{-i\omega t}\sum_{l=0}^{\infty}a_{l}R_{l}(\rho)e^{-l\phi} \label{ansatz}
\end{equation}
where $a_{l}$ are constants. Note that in principle, the sum over $l$ should run from $-\infty$ to $+\infty$, but as hyperbolic materials behave as damping plasmas in the angular direction, whirling waves are always evanescent and therefore $a_l=0$ $\forall l<0$, or equivalently $0 \leqslant l \leqslant +\infty$. Substituting (\ref{ansatz}) into d'Alembert wave equation, one is left with Bessel's equation of order $l/\gamma$: 
\begin{equation}
\rho^2\frac{d^2 R_l}{d\rho^2}+\rho\frac{d R_l}{d\rho}+\left(\rho^2\omega^{2}-\frac{l^2}{\gamma^2}\right)R_{l}=0.
\end{equation}
Solutions to this differential equation are Bessel's functions of fractional order:
\begin{equation}
R_{l}\left(\omega\rho\right)=c_1 J_{l/\gamma}\left(\omega\rho\right)+c_2 J_{-l/\gamma}\left(\omega\rho\right) \label{Req}
\end{equation}
where 
\begin{equation}
J_{l/\gamma}\left(\omega\rho\right)=\sum_{p=0}^{\infty}\frac{(-1)^p}{k!(k+1+l/\gamma)!}\left(\frac{\omega\rho}{2}\right)^{2p+l/\gamma}
\end{equation}
Since Eq. (\ref{Req}) considers both $\pm l/\gamma$ and since $l\geq 0$ we substitute $|\gamma|$ for $\gamma$ and choose $c_2=0$ so that the wave amplitude remains finite at $\rho=0$. This leaves us with 
\begin{equation}
|\Phi\left(t,\rho,\phi\right)|^2=|\sum_{l=0}^{\infty}a_{l}J_{l/|\gamma |}(\omega\rho)e^{-l\phi}|^2,
\end{equation}
where the constant $c_1$ was absorbed into $a_l$. Since $$\lim_{\rho\rightarrow 0} J_{l/|\gamma |}(\omega\rho)=0,$$ as the light whirls around and approaches the origin, its amplitude decreases. 

Provided that light propagates in the half space containing the defect, it will always end whirling around the hyperbolic disclination until it reaches the core: therefore, such defect can be understood as an omnidirectional light absorber (analog of an optical black hole, as stated in \cite{Narimanov}). Besides, contrary to coherent perfect absorbers \cite{Chong}, it does not require incident coherent light to be efficient. However, practical realization sets limits to the efficiency of such device. First, electromagnetic energy accumulated at the core of the defect is converted into thermal internal energy, such that the stability of the director field configuration can be maintained from additional cooling devices. Second, the perfect absorption only occurs within a limited frequency bandwidth due to the resonant nature of the used core-shell spheres. Third, as previous phenomena concerns the extraordinary mode, an efficient optical absorber should include a filter to shut off the ordinary wave. Finally, it should be noticed that the present model concerns optics inside a bulk hyperbolic material: to design a perfect optical absorber, the hyperbolic medium must be impedance matched to avoid sizable reflections at the interfaces.

In this paper, we examined classical optics near a hyperbolic disclination both in the geometrical optics limit and in the wave approximation. Near such defect, light propagation occurs as there were two timelike coordinates in the effective metric, without causing issues related to causality breaking. More importantly, hyperbolic nematics behave like a perfect light absorber in the presence of the defect. They perfectly absorb any incoming radiation and turn it into thermal energy. Such effect is preserved even when incoming light beams are not coherent: despite efficiency bounds related to its practical implementation, this can be used to design a perfect absorber that works even with incoherent light, which to the best of our knowledge, was never considered before. Many modern-day applications could benefit from devices based on such hyperbolic nematics: optical communications, solar energy conversion \cite{Gmachl} or even in medicine (selective delivering of energy in biological tissues for therapy or imaging). As another application of the system studied here we can also think of a metamaterial realization of the two-dimensional Milne universe \cite{gron}. This will be the theme of a separate publication.

\section*{Acknowledgments}

The authors would like to thank the referees for their insightful comments on practical issues related to the design of the perfect optical absorber.  


\global\long\def\theequation{A-\arabic{equation}}
 \setcounter{equation}{0} 


\end{document}